\documentclass[onecolumn]{aastex631}

\usepackage[switch]{lineno}

\usepackage[utf8]{inputenc}
\usepackage{amssymb}
\usepackage{newunicodechar}

\shorttitle{Correlation between dust continuum and CN line emissions in high-mass star-forming regions}
\shortauthors{Hwang et al.}

\begin{document}

\title{The spatial correlation between CN line and dust continuum emitting regions in high-mass star-forming clouds}

\author[0000-0001-7866-2686]{Jihye Hwang}
\email{hjh3772@gmail.com}
\affil{Korea Astronomy and Space Science Institute (KASI), 776 Daedeokdae-ro, Yuseong-gu, Daejeon 34055, Republic of Korea}

\author[0000-0002-3179-6334]{Chang Won Lee}
\affil{Korea Astronomy and Space Science Institute (KASI), 776 Daedeokdae-ro, Yuseong-gu, Daejeon 34055, Republic of Korea}
\affil{University of Science and Technology, Korea (UST), 217 Gajeong-ro, Yuseong-gu, Daejeon 34113, Republic of Korea}

\author[0000-0002-1229-0426]{Jongsoo Kim}
\affil{Korea Astronomy and Space Science Institute (KASI), 776 Daedeokdae-ro, Yuseong-gu, Daejeon 34055, Republic of Korea}

\author[0000-0003-0014-1527]{Eun Jung Chung}
\affil{Department of Astronomy and Space Science, Chungnam National University, 99 Daehak-ro, Yuseong-gu, Daejeon 34134, Republic of Korea}

\author[0000-0002-1229-0426]{Kee-Tae Kim}
\affil{Korea Astronomy and Space Science Institute (KASI), 776 Daedeokdae-ro, Yuseong-gu, Daejeon 34055, Republic of Korea}
\affil{University of Science and Technology, Korea (UST), 217 Gajeong-ro, Yuseong-gu, Daejeon 34113, Republic of Korea}

\begin{abstract}
Measuring the strength of three dimensional (3D) magnetic field vector is challenging as it is not easy to recognize whether its line-of-sight (LOS) and plane-of-sky (POS) components are obtained from the same region. CN ($N = 1 - 0$) emission has been used to get the LOS component of a magnetic field (B$_\mathrm{LOS}$) from its Zeeman splitting lines, while dust continuum emission has been used to get the POS component of a magnetic field (B$_\mathrm{POS}$).  
We use the CN ($N = 1 - 0$) data observed with the Taeduk Radio Astronomy Observatory (TRAO) 14-m telescope and the dust continuum data from $Herschel$ archive toward six high-mass star-forming regions in order to test whether CN line and dust continuum emission can trace a similar region and thus can be used for inferring 3D magnetic field strength.  
Our comparison between CN and H$_2$ column densities for all targets indicates that CN line emission tends to be strong toward bright continuum regions. The positions of peak CN column densities are particularly well correlated with those of peak H$_2$ column densities at least over the H$_2$ column density of 8.0 $\times$ 10$^{22}$ cm$^{-2}$ within one or two telescope beam size in all targets, implying that CN line and dust continuum emitting regions are likely spatially coincident. This enabled us to make the reliable measurement of 3D magnetic field strengths of five targets by taking a vector sum of their B$_\mathrm{LOS}$ and B$_\mathrm{POS}$, helping to decide the magnetical criticality of the targets as supercritical or transcritical.
\end{abstract}

\keywords{}

\section{Introduction} \label{sec:intro}

Magnetic fields play important roles in star-forming processes by supporting star-forming regions against gravitational collapse and regulating core fragmentation (e.g., \citealt{Mouschovias2006, Hwang2021, Hwang2022}, references in \citealt{Pattle2022}). To judge the importance of magnetic field compared with turbulence and gravity, it is necessary to measure their three-dimensional (3D) magnetic field strengths (e.g., \citealt{Mouschovias1976, Crutcher2004}). However, it is a challenging task to directly estimate 3D magnetic field strengths of star-forming regions. 

The plane-of-the-sky (POS) and the line-of-sight (LOS) components of the magnetic fields in star-forming regions can be measured by observations of dust polarized emission (e.g., \citealt{Pattle2022}, and references therein) and Zeeman effect in line emission (e.g., \citealt{Crutcher2019}, and references therein), respectively. 
The measurement of the POS component of a magnetic field is based on the assumption that the minor axis of a dust grain is aligned along a magnetic field line by the Radiative Alignment Torques \citep{Lazarian2007} and the dust continuum is preferentially emitted (polarized) perpendicular to the magnetic field in the POS \citep{Andersson2015}. The magnetic field strength in the POS component has been estimated by the Davis-Chandrasekhar-Fermi (DCF; \citealt{Davis1951, ChanFer1953}) method. In the method, magnetic field lines are assumed to be initially uniform and distorted by non-thermal gas motion, and thus the magnetic field strength can be estimated on a hypothesis that polarization angle dispersion would indicate the distortion of the magnetic field lines by the non-thermal gas motion (e.g., \citealt{Pattle2019, Pattle2022}). 
The LOS component of the magnetic field can be measured by the Zeeman effect of a molecular line tracer whose energy levels are split under the magnetic field. Spectral lines of HI, OH, CN and masers lines of CH$_3$OH and H$_2$O are known to be the tracers showing the Zeeman effects and thus have been used to estimate the LOS magnetic field strengths in star-forming regions (e.g,\citealt{Crutcher2019}).  
In principle, by taking the vector sum of the POS and LOS components, 3D magnetic field strengths in star-forming regions can be obtained, but their practical measurement has been rare and difficult.

There are a few attempts to estimate 3D magnetic field orientations and strengths. \citet{Myers1991} and \citet{Goodman1994} measured the 3D magnetic field strengths in dark cloud complexes by using observations of optical dust polarization and Zeeman effect of {\sc Hi} and OH, which trace diffuse regions ($n_\mathrm{H}$ $\sim$ 10-10$^3$ cm$^{-3}$). The 3D magnetic field in a star-forming region, Barnard 1, was estimated using OH Zeeman effect and dust polarized emission obtained by SCUPOL and POL-2 \citep{Goodman1989, Matthews2002, Coude2019}.
There is one attempt to measure the 3D magnetic field strength in a dense star-forming clump G35.20w in W48 \citep{Pillai2016} with the vector sum of both POS and LOS magnetic field strengths measured by dust polarized emission and CN Zeeman effect. Another attempt measuring 3D magnetic field toward NGC 1333 has been made by combining the results from dust polarization data and toy models assuming a morphology of magnetic field lines perpendicular to filament structures in NGC 1333\citep{Doi2020, Doi2021}.

 One of the important assumptions made in estimating the 3D magnetic field strength using the vector sum is that POS and LOS components of a magnetic field would come from a physically same region. Therefore, it is necessary to check whether dust emission and Zeeman splitting line emission trace the common region. In this paper, we attempt to investigate how dust continuum and CN emissions are correlated with each other towards several star-forming regions. \citet{Hily2008} have shown a tight correlation between 1.3 mm dust and CN emissions. However, the number of their targeted regions was limited to only two low-mass prestellar cores, L1544 and L184. Examining the correlation with the more increased number of samples in various star-forming regions would be necessary in order to increase the measurement reliability of the 3D magnetic field strength obtained from combining POS and LOS components of magnetic fields. 
 
 Here, we present the results of our study for the correlation between CN line and dust continuum emission in six high-mass star-forming regions, W3 OH, OMC 1, NGC 2024, Mon R2, DR21, and S140. We selected the target sources from \citet{Falgarone2008} where CN Zeeman splitting lines were detected at their one or two positions except for Mon R2 where there is no CN Zeeman observation reported yet. In our study, we made CN line observations over all the six targets using the 14-m telescope at the Taeduk Radio Astronomy Observatory (TRAO). We estimated CN and H$_2$ column densities using the CN integrated intensities and $Herschel$ dust emission at 160, 250, 350 and 500 $\mu$m to compare their distribution, finding that there is a fairly good correlation between CN and dust emission. This result ensures a credibility of the  estimation of 3D magnetic field strengths by combining dust polarization and CN Zeeman line observations. In our paper we give the 3D magnetic field strengths for our five high-mass star-forming regions, except for Mon R2.
 
This paper is organized as follows. In section \ref{sec:obs}, we describe how we selected the six target sources and obtained the CN line and dust continuum data towards them. We explain in section \ref{sec:res} how CN and H$_2$ column density maps are constructed. Then we discuss in section \ref{sec:dis} the similarity between the distributions of CN and dust emissions with their peak column density positions, a correlation between CN and H$_2$ column densities, and the 3D magnetic field strengths and their related mass-to-flux ratios in five high mass star-forming regions. Our conclusions are given in section \ref{sec:con}.

\begin{deluxetable*}{ccccccccc}
\tablecaption{Observed targets\label{tab:targets}}
\tablecolumns{9}
\tablenum{1}
\tablewidth{0pt}
\tablehead{(1) & (2) & (3) & (4) & (5) & (6) & (7) & (8) & (9)\\
\colhead{Source name} &
\colhead{R.A.} &
\colhead{Dec.} & \colhead{Mapping size } & \colhead{$v_\mathrm{LSR}$} & \colhead{$T^*_A$} & $\sigma_{\mathrm{CN}}$ & $\Delta$v & $d$
\\ &  (J2000) & (J2000) & [$'\times '$] & [km s$^{-1}$] & [K] & [K km s$^{-1}$] & [km s$^{-1}$, km s$^{-1}$] & [kpc]}
\startdata
W3OH & 02:27:04.1 & 61:52:22 & 10$\times$10 & -48.1 & 0.3 & 0.13 & [-54, -39] & 2.00$^{+0.29}_{-0.23}$ \\
OMC1 & 05:35:14.5 & -05:22:07 & 10$\times$10 & 8.3 & 0.8 & 0.27 & [2, 17] & 0.39 $\pm$ 0.01 \\
NGC2024 & 05:41:44.2 & -01:55:41 &10$\times$10 & 11.0 & 1.1 & 0.31 & [3, 18] & 0.36 $\pm$ 0.02 \\
MonR2 & 06:07:46.0 & -06:22:45 & 10$\times$10 & 9.8 & 0.4 & 0.14 & [3, 17] & 0.78 $\pm$ 0.04 \\
DR21 & 20:38:59.9 & 42:22:38 & 6$\times$6 & 1.8 & -3.6 & 0.5 & [-7, 7] & 1.40$^{+0.12}_{-0.11}$ \\
S140 & 22:19:17.1 & 63:18:35 & 6$\times$6 & -7.2 & 1.5 & 0.5  & [-12, 2] & 0.91 $\pm$ 0.01 \\
\enddata
\tablecomments{Columns: (1) Name of source. (2) and (3) R.A. and Dec. of the central position of CN line OTF map, where the CN Zeeman observations have been made \citep{Falgarone2008}. (4) The OTF mapping size for the TRAO observations. (5) LSR velocity of CN sixth hyperfine component at the OTF mapping center. (6) The peak antenna temperature of the sixth hyperfine component of CN line at the OTF mapping center. (7) The rms value of CN integrated intensity. (8) The velocity range for the integrated intensity map for the sixth hyperfine component of CN line in Fig. 2. (9) The distance of the target. The references are given in Appendix \ref{sec:targets}.}
\end{deluxetable*}

\section{Observations and Data Reductions} \label{sec:obs}

\subsection{Targets}

As observing targets, we selected five massive star-forming regions for which CN Zeeman observations have been made toward one or two positions with the Institut de Radioastronomie Millim\'etrique (IRAM) 30m telescope for measuring the line-of-sight strengths of the magnetic fields (Table \ref{tab:targets};  \citealt{Falgarone2008}). 
We included Mon R2 in our target list although its CN Zeeman effect has not been observed yet, because it is expected to show a strong CN line emission from its strong sub-millimeter continuum emission as a nearby high mass star-forming region \citep{Hwang2022} and can be a good candidate for the future CN Zeeman measurement. 
The information of our observing targets is listed in Table \ref{tab:targets}. The detailed information of each target is described in Appendix \ref{sec:targets}.

\subsection{Observations with TRAO 14-m telescope}

We carried out observations of CN ($N$ = 1$\rightarrow$0) spectral lines towards six high-mass star-forming regions using TRAO 14-m telescope located in Daejeon, South Korea\footnote{\url{https://trao.kasi.re.kr}}, from October to November, 2018 with the on-the-fly (OTF) mapping mode by using a receiver array system, a so-called the SEcond QUabbin Optical Image Array (SEQUOIA) equipped with a 4$\times$4 MMIC preamplifiers. Two IF modules  allow us to observe two sub-bands in the frequency ranges of 85 and 100 GHz, or 100 and 115 GHz, simultaneously. The backend system provides each band to have 4096 channels with a resolution of 15 kHz and a bandwidth of 62.5 MHz. The beam efficiencies are 0.48 and 0.46 at 90 and 110 GHz, respectively \citep{Jeong2019}.

\begin{figure}[ht!]
\epsscale{1.1}
\plotone{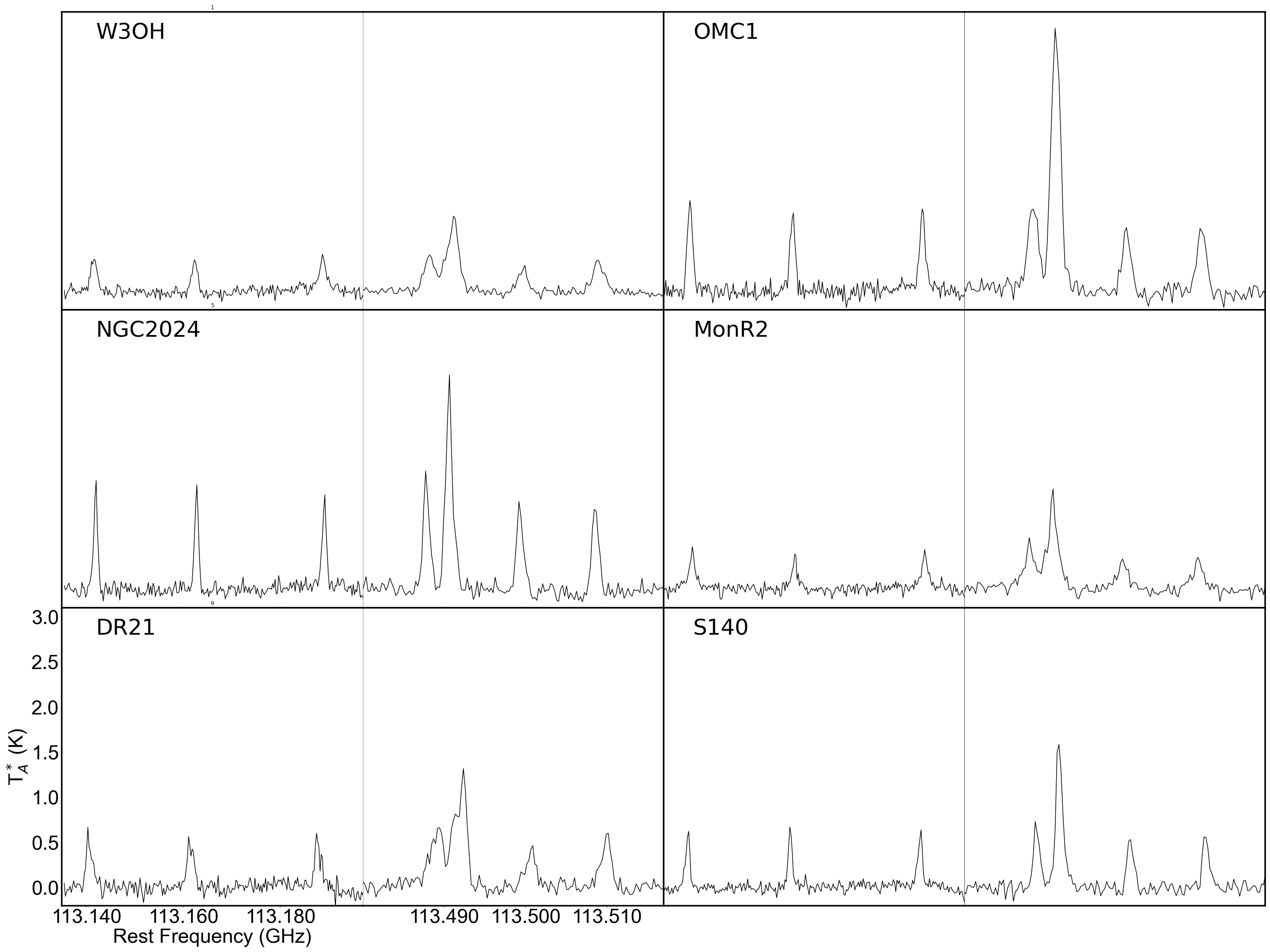}
\caption{CN ($N$ = 1$\rightarrow$0) spectra toward the center positions of the OTF maps for each target (Table \ref{tab:targets}). In each panel a thin solid line divides the seven hyperfine lines into two groups; the one group of three lines with energy level transitions of $N$ = 1 $\rightarrow$ 0, $J$ =1/2 $\rightarrow$ 1/2 and the other group of four lines with the transitions of $J$ =3/2 $\rightarrow$ 1/2. \label{fig:spec}}
\end{figure}

The CN ($N$ = 1$\rightarrow$0) spectral lines have nine hyperfine structures (HFS). Our observations were simultaneously made for seven components out of them (Table \ref{tab:CN}) using two IF modules whose center frequencies are 113.17495 and 113.49097 GHz, respectively. The former and latter frequencies were tuned to observe the first three lines in Table \ref{tab:CN} using the first IF module and the remaining four lines using the second IF module. Mapping observations were made over areas of either  6$' \times$ 6$'$ or 10$' \times$10$'$ in the six high-mass star-forming regions (Table \ref{tab:targets}). The center positions of mapping regions are listed in Table \ref{tab:targets}. The system temperatures vary from 320 to 540 K during our observations. The beam size of the TRAO is 46$''$ at 113 GHz band. Our observing data were reconstructed with the Nyquist sampling of a pixel size of 22$''$ and a spectral resolution of 0.3 km s$^{-1}$. The observation time spent for each source is either 60 or 100 minutes for the mapping size of 6$'$ $\times$ 6$'$ or 10$'$ $\times$ 10$'$, respectively. We made at least four OTF scan observations for each target and then averaged all OTF scans to get its one averaged map. The line data analysis was performed using Continuum and Line Analysis Single-dish Software (CLASS; \citealt{Pety2005}; \citealt{gildasteam2013})\footnote{\url{https://www.iram.fr/IRAMFR/GILDAS}}. 
All seven hyperfine components of CN in the observing frequency range towards six targets were strongly detected  with a high signal-to-noise ratio (S/N) $>$ 10 for the brightest component at their peak intensity positions. The mean brightness temperatures of the sixth hyperfine component at 113.49964 GHz used in our analysis are found to range from 0.17 to 0.54 K in our targets, with the mean rms temperatures of 0.04 $\sim$ 0.08 K. Figure \ref{fig:spec} shows the CN spectra of all targets at the OTF map central positions given in Table \ref{tab:targets}. We smoothed each spectrum in the velocity domain to have its resolution of 0.6 km s$^{-1}$. The peak values of the spectra are shown in Table \ref{tab:targets}.

\begin{figure}[ht!]
\epsscale{1.2}
\plotone{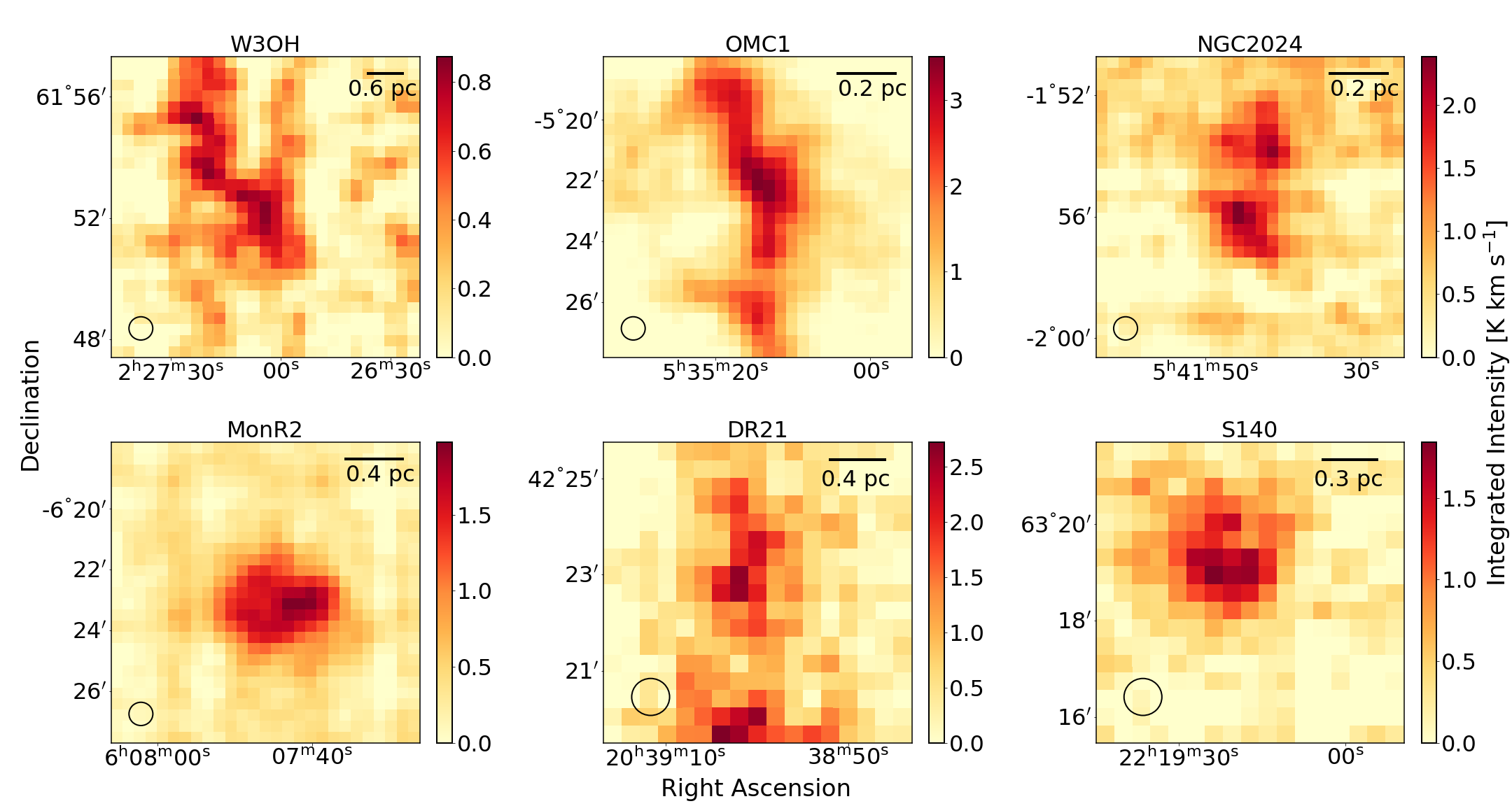}
\caption{Integrated intensity maps of the sixth CN hyperfine transition towards the target sources. A circle in the lower left corner of each panel shows the beam size (46$''$) of TRAO 14-m telescope. 
The velocity range for the intensity integration for each map is listed in Table 1. \label{fig:mom0}}
\end{figure}

Figure \ref{fig:mom0} shows the integrated intensity maps for the sixth CN hyperfine component, as the optically thinnest component of the seven ones, towards the six target sources where the intensity is integrated in the velocity ranges listed in Table \ref{tab:targets}. The temperature in the maps is given in the scale of the main beam temperature (T$_\mathrm{R}$) by considering the beam efficiency (46\%) of TRAO at 110 GHz. 

\begin{deluxetable*}{cccc}
\tablecaption{CN ($N$ = 1$\rightarrow$0) hyperfine transitions \label{tab:CN}}
\tablecolumns{4}
\tablenum{2}
\tablewidth{0pt}
\tablehead{(1) & (2) & (3) & (4)\\
\colhead{Line} &
\colhead{Transition} &
\colhead{Freauency} & \colhead{Relative intensity}\\
\colhead{}&\colhead{$N, J, F$ $\rightarrow$ $N'$, $J'$, $F'$} &\colhead{GHz}
}
\startdata
1 & 1, 1/2, 1/2 $\rightarrow$ 0, 1/2, 3/2 & 113.14416 & 8\\
2 & 1, 1/2, 3/2 $\rightarrow$ 0, 1/2, 1/2 & 113.17049 & 8\\
3 & 1, 1/2, 3/2 $\rightarrow$ 0, 1/2, 3/2 & 113.19128 & 10\\
4 & 1, 3/2, 3/2 $\rightarrow$ 0, 1/2, 1/2 & 113.48812 & 10 \\
5 & 1, 3/2, 5/2 $\rightarrow$ 0, 1/2, 3/2 & 113.49097 & 27\\
6 & 1, 3/2, 1/2 $\rightarrow$ 0, 1/2, 1/2 & 113.49964 & 8\\
7 & 1, 3/2, 3/2 $\rightarrow$ 0, 1/2, 3/2 & 113.50891 & 8 \\
\enddata
\end{deluxetable*}

\subsection{Herschel data}

We used dust continuum data obtained with the $Herschel$ Space Observatory to get H$_2$ column density maps in order to compare them with CN column density maps. The data for all the targets were retrieved from $Herschel$ archive. The $Herschel$ is equipped with two photometric instruments, Photodetector Array Camera and Spectrometer (PACS; \citealt{Poglitsch2010}) and Spectral and Photometric Imaging REceiver (SPIRE; \citealt{Griffin2010}) which enables to observe a field in five photometric bands at 70, 160, 250, 350, and 500 $\mu$m. The beam sizes of the $Herschel$ are 5.5$''$, 11.5$''$,18$''$, 25$''$ and 36$''$  at 70, 160, 250, 350, and 500 $\mu$m, respectively. The continuum images at 160, 250, 350 and 500 $\mu$m for six targets were cropped to have the same sizes as  the ones of CN maps in each target (Table \ref{tab:targets}). The observation Identity Documents (IDs) of the data are 1342189702, 1342218967, 1342215984, 1342204052, 1342186840, and 1342187331 for W3OH, OMC1, NGC2024, MonR2, DR21 and S140, respectively.

\begin{figure}[ht!]
\epsscale{1.2}
\plotone{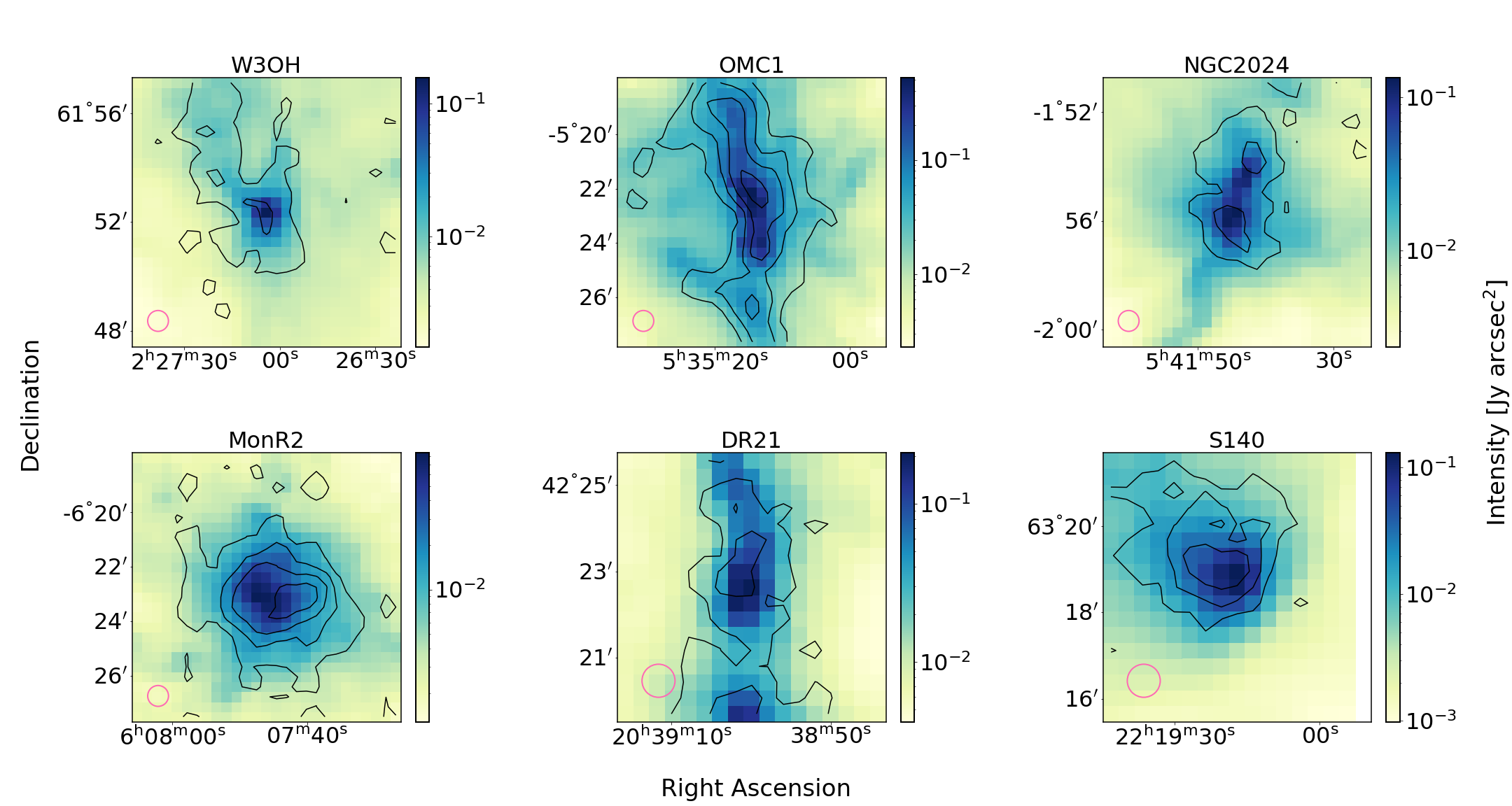}
\caption{Color images of dust continuum obtained with $Herschel$ at 500 $\mu$m. The $Herschel$ images are convolved to the TRAO beam size, 46$''$, which is marked as a pink circle in the lower left corner. Contours indicate the distribution of CN integrated intensities from 3$\sigma_{\mathrm{CN}}$ in steps of 3$\sigma_{\mathrm{CN}}$ where the $\sigma_{\mathrm{CN}}$ is the uncertainty of CN integrated intensity in unit of K km s$^{-1}$ listed in Table \ref{tab:targets}. \label{fig:dust}}
\end{figure}

\subsection{JCMT data}\label{sec:jcmt}

The dust polarization data of the targets were retrieved from the data archive of the James Clerk Maxwell Telescope (JCMT) observatory. The polarization data for all targets except for S140 were obtained from observations by a polarimeter POL-2 \citep{Friberg2016} inserted in the Submillimeter Common-User Bolometer Array 2 (SCUBA-2; \citealt{Holland2013}) camera, while the dust polarization data for S140 were taken using the previous polarimeter the JCMT, the SCUPOL.  
Project codes of the Pol-2 data used in this paper are M16AL004 (OMC-1; P.I. Derek Ward-Thompson), M17BL011 (Mon R2, DR21; P.I. Derek Ward-Thompson), M21BP020 (NGC2024; P.I. Nicolas Peretto), and E21BJ002 (W3OH; P.I. Shu-ichiro Inutsuka). The S140 data were obtained from the SCUPOL legacy catalog \citep{Matthews2009}.

\section{Results} \label{sec:res}

\subsection{Distribution of CN line emission and $Herschel$ continuum emission}
Figure \ref{fig:dust} compares CN line emission in contours with dust continuum emission in color tones for six targets, indicating
that the regions emitting CN emission are well matched with those emitting dust emission. Especially, the peak CN and dust intensities are very well correlated, indicating that the CN and dust emission may be coming from the similarly high density regions. We attempt to examine at which column density level CN line and dust continuum would similarly emit in the dense regions. For this purpose, we estimate CN column density using our CN observations and H$_2$ column density using the $Herschel$ data in the next two sections and examine their distribution features in the discussion section.

\subsection{CN column density}

\begin{figure}[ht!]
\epsscale{1.1}
\plotone{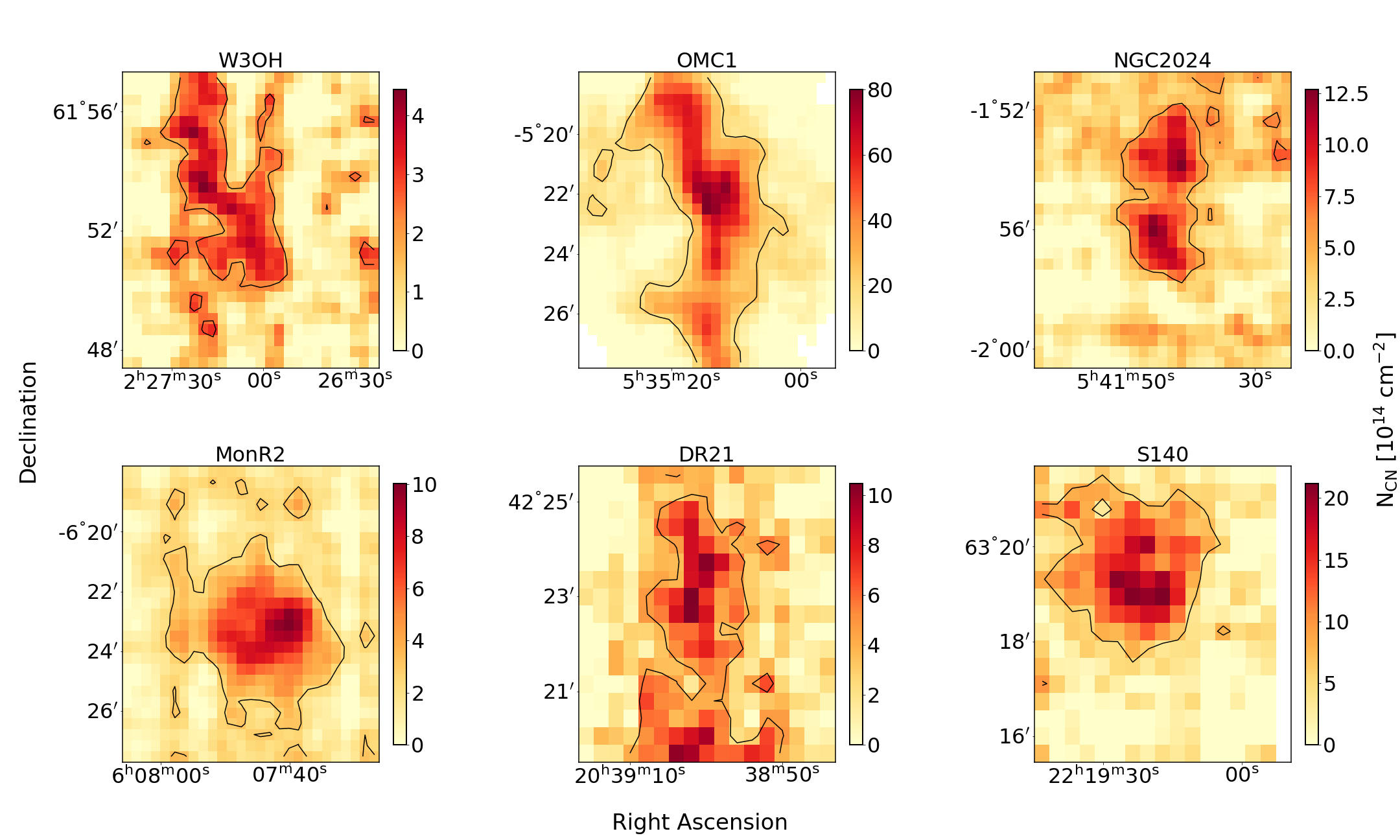}
\caption{Maps of CN column density of the target sources.  
Black contours in each panel delineate a 3$\sigma_\mathrm{CN}$ level.\label{fig:CNcol}}
\end{figure}

The CN column density can be estimated  from the following equation derived with optically thin and the Rayleigh-Jeans approximation ($h \nu \, \ll k T_\mathrm{ex}$ at 113 GHz) (\citealt{Mangum2015}),

\begin{equation}
N_\mathrm{CN}=\bigg(\frac{3k}{8\pi^3 S\mu_0^2R_i}\bigg)\bigg(\frac{Q_\mathrm{rot}}{g_Jg_Kg_I}\bigg) \exp\bigg( \frac{E_u}{kT_\mathrm{ex}} \bigg) \int  \frac{T_R T_\mathrm{ex}}{f(T_\mathrm{ex}-T_\mathrm{bg})} d\upsilon, \label{eq:cn}
\end{equation}

\noindent
where $k$ is the Boltzmann constant, $S$ is the line strength, $\mu_0$ is the dipole moment, $R_i$ is the relative strength of all transitions, $Q_\mathrm{rot} = kT_\mathrm{ex}/h B_0 + 1/3$ is the rotational partition function that represents a statistical sum over all rotational energy levels in the molecule, $h$ is Planck's constant, $B_0$ is the rigid rotor rotation constant, $T_\mathrm{ex}$ is the excitation temperature, $E_u$ is the energy in upper level $u$, 
$g_J$ is the rotational degeneracy, $g_K$ is the $K$ degeneracy associated with the internal quantum number $K$, $g_I$ is the nuclear spin degeneracy,  $T_\mathrm{bg}$ is the background temperature assumed to be 2.7 K as the cosmic microwave background temperature \citep{Fixsen2009}, $f$ is the filling factor assumed as a unit, and $\int T_R d\upsilon$ is the integrated intensity of the CN emission. 
For the estimation of the CN column density we used the sixth hyperfine component of CN ($N$, $J$, $F$ $\rightarrow$ $N'$, $J'$, $F'$ = 1, 3/2, 1/2 $\rightarrow$ 0, 1/2, 1/2) which has the least optical depth among the seven HFS components.
We used $\mu_0=1.45$ Debye, $E_u=5.45$ K and $B_0=56693.47$ MHz given by the Jet Propulsion Laboratory (JPL) molecular spectroscopy database\footnote{\url{https://spec.jpl.nasa.gov/ftp/pub/catalog/catdir.html}} and spectral line catalog \citep{Pickett1998}. 
The $Q_\mathrm{rot}$ is given as $0.37 (T_\mathrm{ex}+0.9)$ for the sixth transition of CN. The derivation of the excitation temperature using CN hyperfine lines is found to be not straightforward, usually suffering from a large degeneracy in its determined values especially in the optically thin case of the CN line. Therefore we used the dust temperature as the excitation temperature by assuming that the CN and dust are well mixed. In fact the Figure 3 shows that distributions of CN and dust emission are well matched at least around their emission peak regions and thus using the dust temperatures toward the emission peak regions are thought to be viable.  Figure \ref{fig:CNcol} shows the maps of the estimated CN column density. 
A more detailed comparison between CN and H$_2$ column densities in each region is given in Section \ref{sec:corr}.

\subsection{H$_2$ Column density}\label{subsec:sed}

\begin{figure}[ht!]
\epsscale{1.1}
\plotone{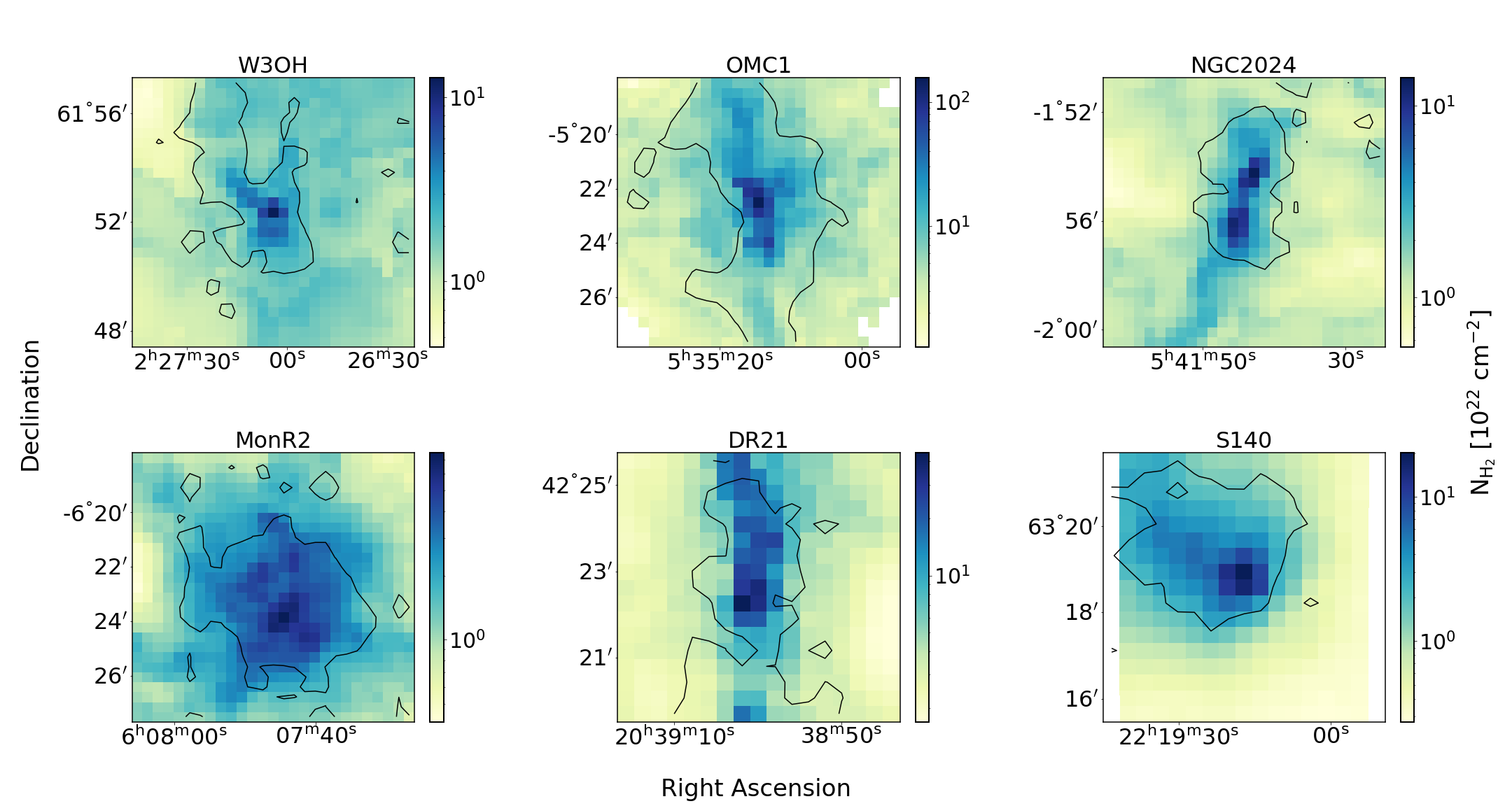}
\caption{Maps of H$_2$ column density of the target sources. The column density is estimated by the SED fitting procedure of the $Herschel$ data at four wavelengths. Black contours indicate the 3$\sigma_\mathrm{CN}$ level. \label{fig:H2col}}
\end{figure}

The H$_2$ column density can be estimated by combining the $Herschel$ 160, 250, 350, and 500 $\mu$m data. The Spectral Energy Distribution of the $Herschel$ data is used to estimate the dust temperature and the 500 $\mu$m data are used to derive the H$_2$ column density with an assumption of a gas to dust mass ratio of 100. The equation for the estimation of the H$_2$ column density is given as follows;  

\begin{equation}
    N_{\text{H}_2} =\frac{I_{\nu}}{\mu m_{\text{H}}  \kappa(\nu) B_\nu(T)},
\label{eq:bb}
\end{equation}

\noindent
where $I_\nu$ is the intensity at the frequency $\nu$ \citep{Hildebrand1983}, $B_\nu$($T$) is the Planck function at dust temperature $T$, $\mu=2.8$ is the mean molecular weight per hydrogen molecule \citep{Kirk2013}, $m_\mathrm{H}$ is the mass of a hydrogen atom, $\kappa(\nu)=\kappa_{\nu_0}(\nu/\nu_0)^\beta$ is the dust opacity function where $\kappa_{\nu_0}$ = 0.1 cm$^2$ g$^{-1}$ at $\nu_0$ = 1000 GHz (\citealt{MotteAndr2001}; \citealt{Andre2010}), and $\beta = 2$ is the dust emissivity index \citep{Beckwith1990}.

We determined the dust temperature and H$_2$ column density for the targets using the above equation through an iterative SED fitting of the $Herschel$ data at 160, 250, 350, and 500 $\mu$m starting from an initial guess of the column density as 10$^{22}$ cm$^{-2}$. 
Note that before the SED fitting procedure, we subtracted the background emission at regions about 9-10$'$ away from the central position of each map to remove any possible emission contribution from the Galactic plane in the targets. Then, we convolved the 160, 250, and 350 $\mu$m data in a different beam size to have a beam size of 500 $\mu$m data, 36$''$, and aligned data pixels at 160, 250, and 350 $\mu$m to the coordinates for 500 $\mu$m data. Then we obtain the best-fit dust temperature giving the lowest $\chi^2$ value from the SED fit using a python package of $curve\_fit$ and estimate H$_2$ column density with this dust temperature using 500 $\mu$m data. The estimated column density is again used for the next round of the SED fit to determine an improved dust temperature, with which the column density is again estimated. The fitting range of temperature was assigned from 2.7 to 100 K. This iteration of the SED fit continues until the re-estimated H$_2$ column density agrees within 1\% of its previous value.  
We chose 500 $\mu$m data in the fit and in the calculation of the H2 column density. This is mainly because the beam size of the 500 $\mu$m data is closer to the beam size of the TRAO data (CN line data) than those for 350 and 250 $\mu$m data. We have also used dust continuum data at 350 and 250 $\mu$m data in deriving the H2 column densities, finding the derived H$_2$ column densities are consistent with the values obtained using 500 $\mu$m data within 5\% of the values from 500 $\mu$m data in their differences. 

From this way we determined the H$_2$ column densities and dust temperatures in all pixels of all targets. The obtained H$_2$ column density maps are shown in Figure \ref{fig:H2col}. In the case of S140, there are no PACS observations at 160 $\mu$m, so we used the data at 250, 350, and 500 $\mu$m for the SED fitting.  
In SPIRE observations, bright sources can be saturated, and so these sources are usually set to be observed in the bright mode \footnote{\url{http://herschel.esac.esa.int/Docs/SPIRE/pdf/spire_om_v24.pdf}}. However, S140 has not been observed with this bright mode. 
In a few central pixels of S140, there are no values at 250 $\mu$m because of the intensity saturation issue. In this case, we used the highest value at 250 $\mu$m at a pixel nearby from those pixels for the SED fitting. In the case of OMC-1, a few pixels of dust continuum at 160 $\mu$m are also found to be saturated, having no values assigned. In this case, we used the data at 250, 350, and 500 $\mu$m data for the SED fitting.

\section{Discussions} \label{sec:dis}

This section discusses how CN and dust emitting regions are spatially related. For this purpose, we examine how peak positions and half maximum (HM) contours are coincident and how well CN and dust emitting regions are physically related. Here the HM contours are meant to be the contours having half values of the peak CN and H$_2$ column densities in their column density distribution. From the likely spatial coincidence between CN and dust emissions, the possible 3D magnetic field strengths are estimated in five high mass star-forming regions and how their measurements would affect on the physical status of the star-forming regions is discussed. For the direct comparison of two column densities, the H$_2$ column density maps derived with $Herschel$ continuum emission were convolved to the TRAO beam size and re-gridded to fit the pixels of the CN column density maps.

\subsection{Spatial relation between CN line and dust continuum emitting regions}

CN line emission has been detected in star-forming regions with densities ranging from 10$^4$ to 10$^6$ cm$^{-3}$ \citep{Turner1975} where there are the dense cores which may start to collapse to form stars. 
However, it has been questionable whether the CN emission showing the Zeeman effect comes from the central regions of the dense cores that can be well traced with dust continuum emission.
Therefore, it is necessary to make a systematic comparison between the peak positions of CN and dust emission.

\begin{figure}[htb!]
\epsscale{1.2}
\plotone{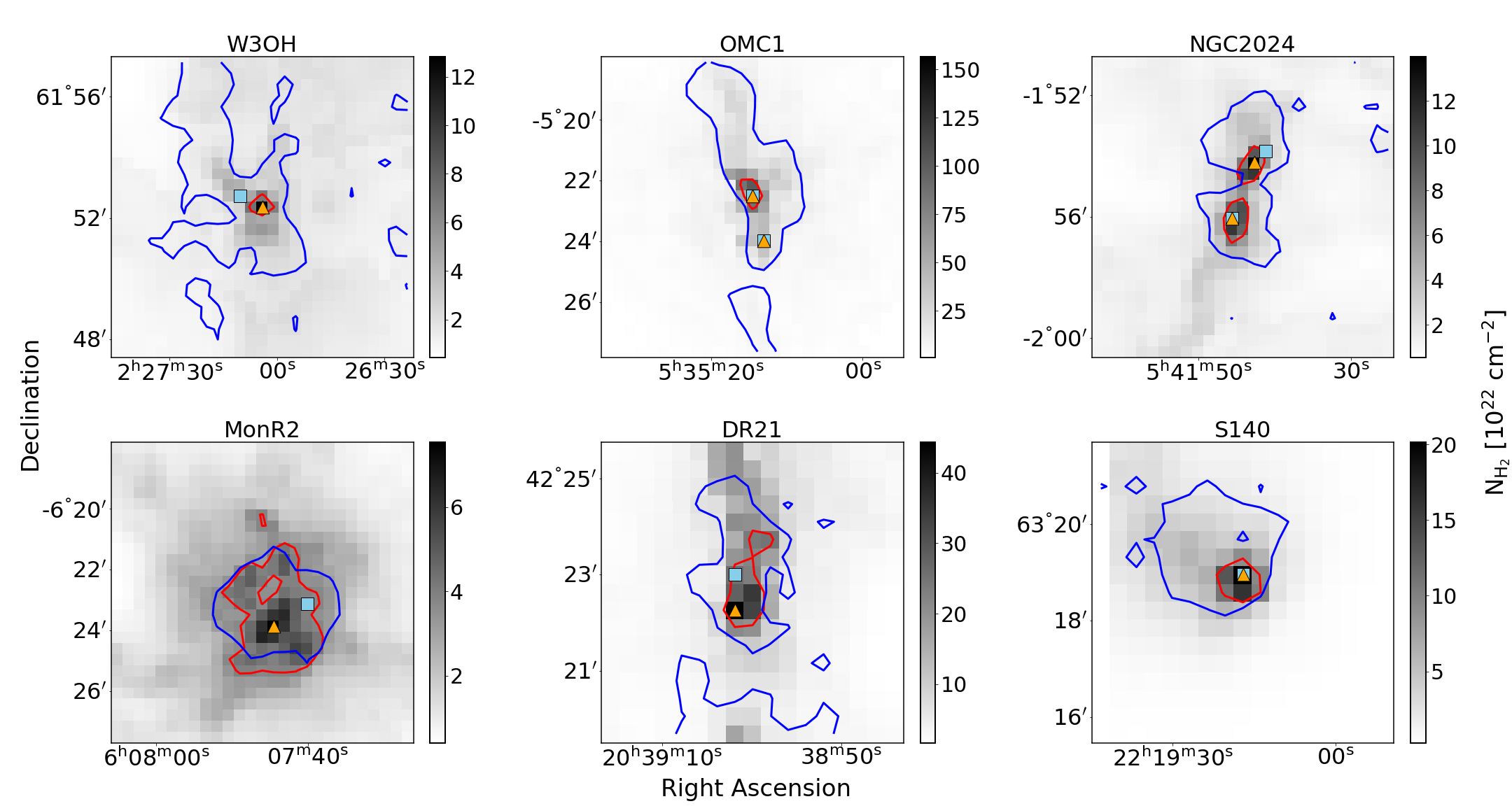}
\caption{The spatial distribution of the CN and H$_2$ column densities. A gray image displays the distribution of the H$_2$ column density. The scale bar in each panel indicates the value of the column density. Blue and red contours in each panel delineate the half maximum levels of CN and H$_2$ column densities, respectively. Cyan squares and orange triangles indicate the peak positions for CN and H$_2$ column densities, respectively. OMC 1 and NGC2024 show two sets of symbols for local column density peaks for two clumps for each source. 
\label{fig:peak_pos}}
\end{figure}

\begin{figure}[htb!]
\epsscale{1.2}
\plotone{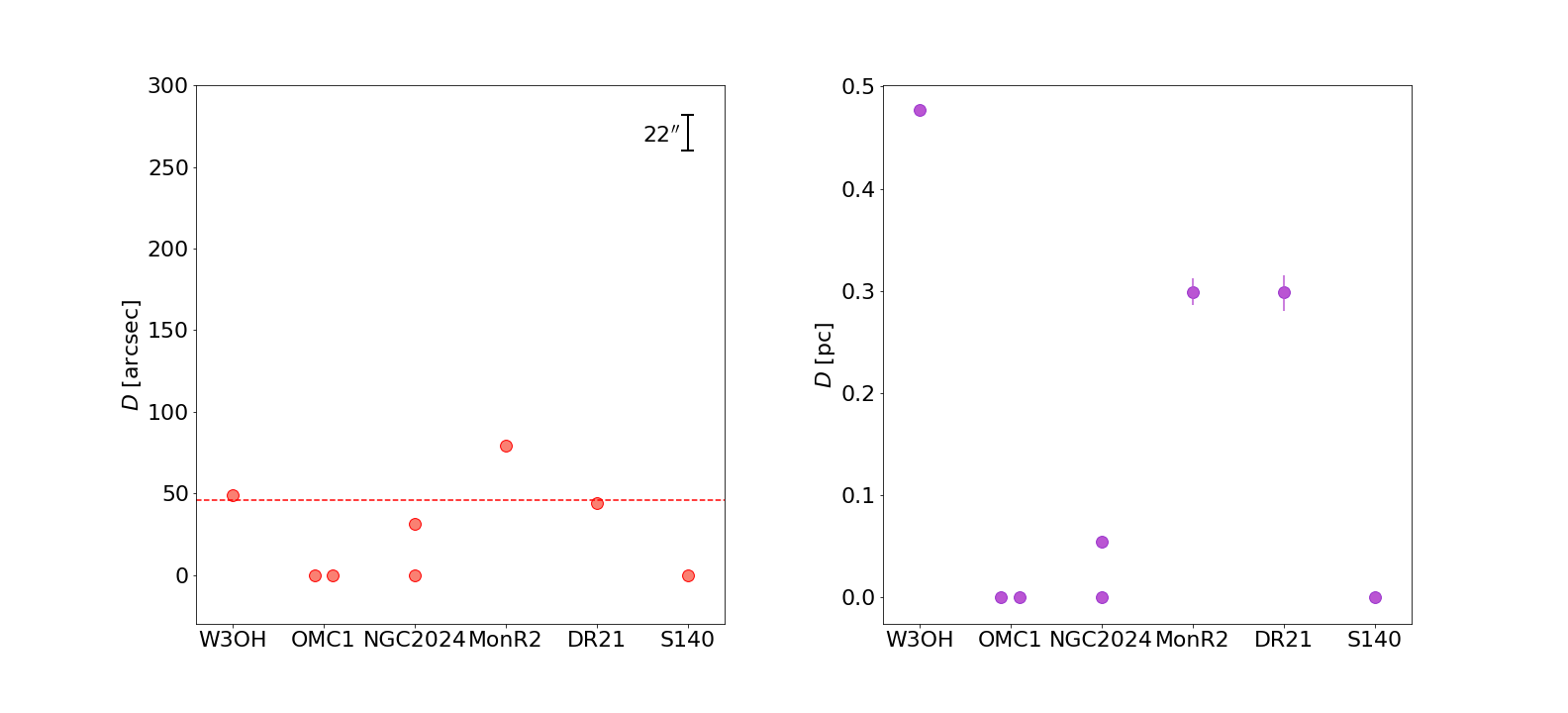}
\caption{The spatial separations between peak positions in CN and H$_2$ column density distributions for the observing targets, in units of arcsec (left) and pc (right). The dashed red line in the left panel indicates the TRAO beam size of 46$''$. A bar for the pixel size of 22$''$ is shown in the right upper corner of the left panel. The error bars shown in the right panel are estimated from the distance uncertainties of the target sources listed in Table \ref{tab:targets}. 
\label{fig:peak_diff}}
\end{figure}

Here we make a similar comparison for CN and H$_2$ column density maps for the six high mass star-forming regions in Figure \ref{fig:peak_pos}. The main purpose of the comparison is to examine how the CN line bright regions are closely related to dust continuum bright regions and thus we added two more peak intensity positions in OMC-1 and NGC2024 showing multiple local peak intensity positions in CN and dust intensity distribution. Therefore we were able to consider 8 strong intensity positions in comparing the peak positions CN line and dust continuum emitting regions, finding that the peak positions of CN column densities in the targets are closely located to the peak positions of H$_2$ column densities. 
The separations of the peak positions are within or less than the beam size of the TRAO, $\sim 46''$, except for Mon R2 whose separation is $\sim 79''$ (see Figure \ref{fig:peak_diff}).

We also examine the features of CN line and dust emission distributions by comparing the half maximum (HM) contour areas of CN and H$_2$ column density maps (Figure \ref{fig:peak_pos}). We find that the HM contour areas of CN column density maps are usually wider, by at least 7 times, than those of the H$_2$ column density maps in all targets, except for Mon R2 where such areas of CN and H$_2$ column densities are comparable. However, we note that the areas of $>$ 5$\sigma$ level are wider in the H$_2$ column density maps than CN column density  maps because of the superior sensitivity in dust continuum data (Appendix \ref{sec:area}), indicating that CN emission is more widely distributed in the dense core regions than the dust continuum emission while dust continuum emission traces more extended area in the low density regions. 

In summary CN emission is usually bright at the dust continuum bright region, tracing more extended area than the dust emitting area in the dense core regions. Dust continuum data are more sensitive enough to trace the low density extended area than CN emission data.

\subsection{Correlation between CN and H$_2$ column densities}\label{sec:corr}
\begin{figure}[htb!]
\epsscale{1.1}
\plotone{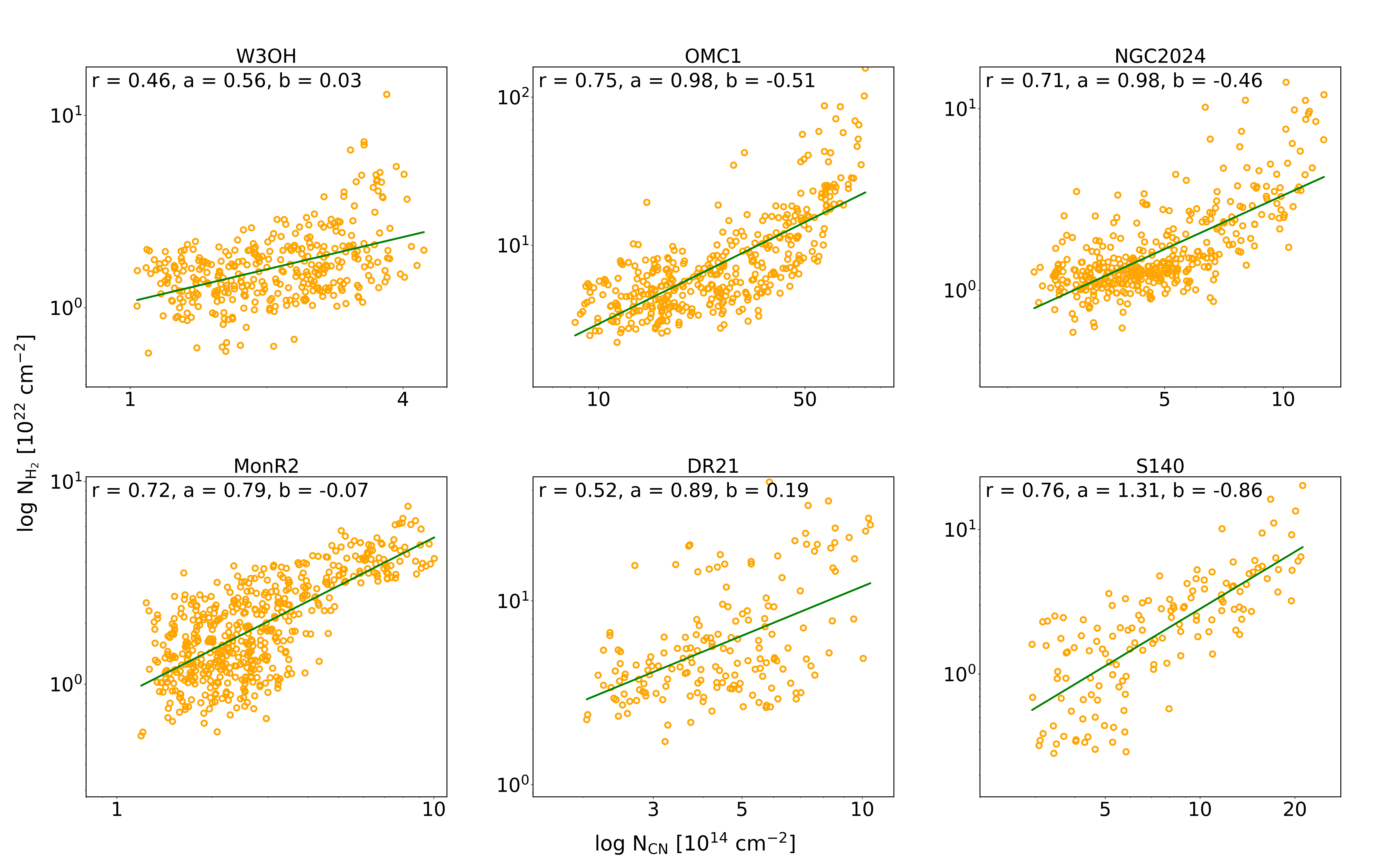}
\caption{Correlation between column densities of CN and dust toward six target sources. A green line in each panel draws a best fitting result of both column densities with a function, log $N_\mathrm{d}$ = $a$ log $N_\mathrm{CN}$ + $b$. The correlation coefficient, $r$, and the fitting parameters, $a$ and $b$, are given at the upper part of each panel.  
\label{fig:col}}
\end{figure}

How CN line emitting regions do relate to the dust continuum emitting regions can be also examined with the correlation between CN and H$_2$ column densities  toward the targets as shown in Figure \ref{fig:col}. The figure plots two column densities within regions enclosed by 3$\sigma_\mathrm{CN}$ level (Figure \ref{fig:CNcol}) for which the H$_2$ column densities are at least larger than 1.6 $\times$ 10$^{21}$ cm$^{-2}$. The correlation coefficients $r$ between both column densities are larger than 0.7 in four targets, indicating a good correlation between two column densities in those sources. In case of two targets, W3OH and DR21, the correlation coefficients are 0.46 and 0.52, respectively, meaning that the correlation between two column densities in two targets is not as tight as those for other four targets. Even in this case, however, we note that two CN and H$_2$ column densities are positively correlated in the sense that the CN column densities get higher at the high H$_2$ column density regions.

\subsection{3D magnetic field strength}
In previous sections we found that CN column densities are positively correlated with H$_2$ column densities in the targets. In particular, the spatial positions showing the peak column density of CN are even better correlated with the positions showing the peak H$_2$ column density within one or two TRAO beam size in all targets, while the CN emitting area is more extended than the dust emitting area in the dense core regions. From this we see a reliability of combining both LOS and POS components of magnetic field strengths ($B_{\mathrm{LOS}}$ and $B_{\mathrm{POS}}$) measured in the high dense regions to obtain 3D magnetic field strengths of our targets as these two CN line and dust continuum emissions are likely coming from a spatially similar area especially in the dense central regions of the clouds with at least over the H$_2$ column density of 8.0 $\times$ 10$^{22}$ cm$^{-2}$.
Therefore we attempt to estimate the 3D magnetic field strengths towards our targets by making a vector sum of two LOS and POS components of the magnetic fields that have been measured for these targets and discuss the role of the magnetic field in the star formation activity in the targets using their 3D magnetic field strengths. The $B_{\mathrm{LOS}}$ in Mon R2 has not been obtained, and thus its 3D magnetic field strength could not be estimated.

The LOS magnetic field strengths have been obtained by the CN Zeeman effect toward a single position of the targets using the IRAM 30 m telescope with a beam size of 24$''$ \citep{Falgarone2008}. However, the POS magnetic field strengths for each target were not necessarily estimated toward the position where the LOS component is measured, but usually given as mean values over the targets (see  Appendix \ref{sec:targets}). Therefore we re-estimated the POS magnetic field strengths in each target toward the positions where the LOS magnetic field strengths were measured, using the following equation \citep{Crutcher2004};  
\begin{equation}
B_{\text{POS}} \approx 9.3\sqrt{n(\text{H}_2) \, [cm ^{-3}]}\frac{\Delta V \, [km s^{-1}]}{\sigma_\theta \, [\text{degree}]},
\label{eq:cf}
\end{equation}
where $n_{\mathrm{H}_2}$ is the volume density of molecular hydrogen, $\Delta V = 2\sqrt{2\ln2}\sigma_v$ is the full width at half maximum of the non-thermal component of a spectral line, $\sigma_v$ is the velocity dispersion of the non-thermal gas component, and $\sigma_\theta$ is the polarization angle dispersion. 
The volume density was estimated by dividing the column density of H$_2$ width a cloud depth which is given here as an effective radius of $R_\mathrm{eff} = \sqrt{A/\pi}$ where $A$ is the area within the HM contour of the H$_2$ column density map. The estimated H$_2$ volume densities for six targets are listed in Table \ref{tab:dcf}, which are consistent with those determined from other line observations for the same targets within a few factors \citep{Falgarone2008}.
The $\sigma_v$ was estimated from the CN line profile obtained from our TRAO observations using  $\sigma_v$=$\sqrt{\sigma_{obs}^2-kT_k/m_{\text{CN}}}$, where $\sigma_{obs}$ is obtained by the Gaussian fit for the sixth hyperfine component of CN, $T_k$ is the kinetic temperature which is assumed to be the same as the dust temperature, and $m_{\text{CN}}$ is the mass of CN molecule. 

Polarization angle dispersion  ($\sigma_\theta$) was derived 
by using a new technique in the DCF method by \citet{Hwang2021} in which field strengths are measured in a moving box with a small size where any curvature of magnetic field line would not affect the estimation of the field strength. In our measurement of polarization angle dispersion for deriving the POS component of the magnetic field strength, the box size was chosen as small as possible to best fit to the beam size of 24$''$  with which the LOS component was measured. 
By using dust polarization maps obtained by the POL-2 or SCUPOL on the JCMT \citep{Matthews2009, Hwang2021, Hwang2022}, we binned the polarization segments with a pixel size of 12$''$, chose a box of 3 $\times$ 3 pixels containing nine polarization segments, and re-estimated polarization angle dispersion in the box. It turned out that a box size of 3 $\times$ 3 pixels corresponding to the beam size of 36$''$ is significantly smaller than the radius of magnetic field curvature, making the polarization angle dispersion to be smaller than 25$^\circ$ in most case and thus the measurement of POS strength of magnetic field to be reliable. 
The POS magnetic field strengths and other physical parameters used for the DCF method for each source are listed in Table \ref{tab:mag}.

We simply took vector sums of the POS and LOS components of five target sources to estimate their 3D magnetic field strengths ($B$; Table \ref{tab:mag}) which range from 0.6 to 4.9 mG and thus drove the mass-to-flux ratios for our targets that can be useful for discussing on the relative importance of magnetic fields in their star-forming activities.
A dimensionless mass-to-flux ratio ($\lambda$) is given by dividing its observed value by the critical value in a magnetized disk structure
\citep{Nakano1978, Crutcher2004},

\begin{equation}
\lambda=7.6\times10^{-21} \frac{N_{\mathrm{H}_{2}} \, [\mathrm{cm}^{-2}]}{B \, [\mu \mathrm{G}]}.
\end{equation}
 This ratio was estimated using the H$_2$ column density in Table \ref{tab:dcf} and $B$ in Table \ref{tab:mag}. A source for which $\lambda$ is smaller than 1 is referred to be magnetically subcritical, implying that the magnetic field over the source is strong enough against its gravitational collapse, while the source for which $\lambda$ is larger than 1 is magnetically supercritical and in this case the magnetic field is not strong enough to resist against the gravitational collapse over the source.

For making more reasonable estimation for the 3D magnetic field strength and the mass-to-flux ratio, here we discuss possible statistical over-estimation of the $B_{\mathrm{POS}}$ in comparison with the $B_{\mathrm{LOS}}$. 
If 3D magnetic field vectors are randomly oriented, the angle-averaged mean POS and LOS components would be factors of $\pi/4 \sim 0.79$ and 0.5 smaller than the 3D magnetic field strength, respectively \citep{Pillai2016}, and thus the strength of POS component of any 3D magnetic field vector could be statistically $\pi/2 (= 1.58)$ times larger than that of its LOS component. However, \citet{Pattle2022} showed that for all possible compilations of the observed sources, the POS magnetic field strengths measured by the DCF method are averagely 6.3 $\pm$ 1.5 times larger than those obtained by Zeeman measurements. Then it is possible that the measurement of POS strength by the DCF method is statistically overestimated by a factor of $6.3/1.58 \sim4$.
Because of this, we corrected the POS magnetic field strength by dividing a factor of 4 to calculate the corrected 3D magnetic field strength ($B_\mathrm{cor}$) for our targets.
The mass-to-flux ratios ($\lambda_\mathrm{cor}$) of our targets were estimated using this corrected 3D magnetic field strength, which range from 0.9 to 5.2. The estimated $\lambda_\mathrm{cor}$ values for our targets are all listed in Table \ref{tab:mag}, indicating that our sources are magnetically supercritical, except for W3OH which is more likely in a magnetically transcritical status.

\begin{deluxetable*}{cccccc}
\tablecaption{Parameters for estimating the POS component of the magnetic field by the DCF method\label{tab:dcf}}
\tablecolumns{6}
\tablenum{3}
\tablewidth{0pt}
\tablehead{(1) & (2) & (3) & (4) & (5) & (6)\\
\colhead{Source} &
\colhead{$\sigma_\theta$ } &
\colhead{$\sigma_v$} & \colhead{N$_{\mathrm{H}_2}$} & \colhead{$R_\mathrm{eff}$} & \colhead{$n_{\mathrm{H}_2}$}\\
 & [degree] & [km s$^{-1}$] & [10$^{22}$ cm$^{-2}$] & [pc] & [10$^4$ cm$^{-3}$]}
\startdata
W3OH & 39 $\pm$ 2.9 & 1.5 $\pm$ 0.6 & 12.8 $\pm$ 0.2 & 0.24 & 8.6 $\pm$ 0.2 \\
OMC-1 & 10.2 $\pm$ 0.1 & 1.3 $\pm$ 0.7 & 87.3 $\pm$ 8.7 & 0.05 & 302.9 $\pm$ 30.0 \\
NGC 2024 & 19.9 $\pm$ 1.6 & 1.1 $\pm$ 0.3 & 9.9 $\pm$ 0.2 & 0.07 & 21.4 $\pm$ 0.3 \\
Mon R2 & 21.7 $\pm$ 1.9 & 1.9 $\pm$ 0.7 & 3.6 $\pm$ 0.1 & 0.39 & 1.5 $\pm$ 0.03\\
DR21 & 5.9 $\pm$ 0.3 & 1.7 $\pm$ 0.7 & 35.0 $\pm$ 0.7 & 0.24 & 23.8 $\pm$ 0.6 \\
S140 & 10.6 $\pm$ 7.5 & 1.0 $\pm$ 0.4 & 16.2 $\pm$ 1.0 & 0.12 & 21.5 $\pm$ 1.3 \\
\enddata
\tablecomments{Columns: (1) Source name. (2) Polarization angle dispersion obtained using SCUPOL and POL-2 observations from the JCMT (see Section \ref{sec:jcmt}).  (3) Velocity dispersion of CN sixth hyperfine line obtained at the position where CN Zeeman effect has been detected. 
(4) H$_2$ column density. (5) Effective radius for the area within the HM contour of H$_2$ column density map. (6) Volume density obtained by dividing the column density with twice effective radius.}
\end{deluxetable*}

\begin{deluxetable*}{cccccccc}
\tablecaption{Magnetic field strengths and Mass-to-flux ratios\label{tab:mag}}
\tablecolumns{8}
\tablenum{4}
\tablewidth{0pt}
\tablehead{(1) & (2) & (3) & (4) & (5) & (6) & (7) & (8) \\
\colhead{Source} &
\colhead{$B_\mathrm{LOS}$ } &
\colhead{$B_\mathrm{POS}$} & \colhead{$B_\mathrm{POS,cor}$} & \colhead{$B$} & \colhead{$B_\mathrm{cor}$} & \colhead{$\lambda$} & \colhead{$\lambda_\mathrm{cor}$} \\
 & [mG] & [mG] & [mG] & [mG] & [mG] & & }
\startdata
W3OH & 1.10 $\pm$ 0.33 & 0.3 $\pm$ 0.1 & 0.08 $\pm$ 0.03 & 1.1 $\pm$ 0.3 & 1.10 $\pm$ 0.33 & 0.9 $\pm$ 0.2 & 0.9 $\pm$ 0.3\\
OMC-1 & -0.36 $\pm$ 0.08 & 4.9 $\pm$ 2.6 & 1.23 $\pm$ 0.65 & 4.9 $\pm$ 2.6 & 1.28 $\pm$ 0.62 & 1.4 $\pm$ 0.7 & 5.2 $\pm$ 2.6 \\
NGC 2024 & 0.01 $\pm$ 0.12 & 0.6 $\pm$ 0.2 & 0.11 $\pm$ 0.05 & 0.6 $\pm$ 0.2 & 0.15 $\pm$ 0.05 & 1.2 $\pm$ 0.4 & 5.0 $\pm$ 1.7 \\
Mon R2 &  & 0.6 $\pm$ 0.2 & 0.05 $\pm$ 0.03 & & & &\\
DR21 & -0.54 $\pm$ 0.11 & 3.1 $\pm$ 1.3 & 0.78 $\pm$ 0.33 & 3.1 $\pm$ 1.3 & 0.94 $\pm$ 0.29 & 0.8 $\pm$ 0.3 & 2.8 $\pm$ 0.8 \\
S140 & -0.25 $\pm$ 0.09 & 1.0 $\pm$ 0.8 & 0.25 $\pm$ 0.20 & 1.0 $\pm$ 0.8 & 0.35 $\pm$ 0.16 & 1.2 $\pm$ 0.9 & 3.5 $\pm$ 1.5\\
\enddata
\tablecomments{Columns: (1) Source name. (2) LOS magnetic field strength obtained with CN Zeeman observations \citep{Falgarone2008}. In the case of DR21, a mean value of two measurements is given. (3) POS magnetic field strength by the DCF method obtained using parameters listed in Table \ref{tab:dcf}. (4) POS magnetic field strength reduced by a statistical factor of 4. 
 (5) 3D magnetic field strength by a vector sum of two values in columns (2) and (3). (6) 3D magnetic field strength by a vector sum of the POS and the LOS values in columns (2) and (4). (7) Mass-to-flux ratio by the values in column (5) and H$_2$ column density.. (8) Mass-to-flux ratio by the values in column (6) and H$_2$  column density.}
\end{deluxetable*}

\section{Summary}\label{sec:con}

We carried out the observations of the CN $N$ = 1 - 0 hyperfine transitions toward the six high-mass star-forming regions using the TRAO 14-m telescope to examine how the CN emitting regions are coincident with the dust emitting regions.

First of all, CN column densities are found to be positively correlated with H$_2$ column densities all over the targets and in particular the peak CN column density areas are even better correlated with the peak H$_2$ column density areas of the H$_2$ column density of $>$ 8.0 $\times$ 10$^{22}$ cm$^{-2}$ within one or two telescope beam size in all targets, indicating that CN line and dust continuum emitting regions are likely spatially coincident.  This justifies making a vector sum of both LOS components (from the CN Zeeman effect observations)  and POS components (from the dust polarization observations) of magnetic field strengths to obtain 3D magnetic field strengths toward the high column density regions of high-mass star-forming clouds. 

By using this 3D magnetic field strength, the mass-to-flux ratios were estimated to examine the stability of our high-mass star-forming clouds, finding that the sources are magnetically supercritical or transcritical and thus the magnetic field strengths in the sources may not be sufficient to support them against gravitational collapse.

\software{Starlink \citep{Jenness2013}, GILDAS/CLASS (\citealt{Pety2005}; \citealt{gildasteam2013}) }

\begin{acknowledgements}
This research was supported by the Korea Astronomy and Space Science Institute (KASI) under the Research and Development (R\&D) program supervised by the Ministry of Science and ICT.
The JCMT is operated by the East Asian Observatory on behalf of The National Astronomical Observatory of Japan; Academia Sinica Institute of Astronomy and Astrophysics; the Korea Astronomy and Space Science Institute; the Operation, Maintenance and Upgrading Fund for Astronomical Telescopes and Facility Instruments, budgeted from the Ministry of Finance of China. Additional funding support is provided by the Science and Technology Facilities Council of the United Kingdom and participating universities and organizations in the United Kingdom, Canada and Ireland. Additional funds for the construction of SCUBA-2 were provided by the Canada Foundation for Innovation.
The Herschel spacecraft was designed, built, tested, and launched under a contract to ESA managed by the Herschel/Planck Project team by an industrial consortium under the overall responsibility of the prime contractor Thales Alenia Space (Cannes), and including Astrium (Friedrichshafen) responsible for the payload module and for system testing at spacecraft level, Thales Alenia Space (Turin) responsible for the service module, and Astrium (Toulouse) responsible for the telescope, with in excess of a hundred subcontractors.
C.W.L. is supported by the Basic Science Research Program through the NRF funded by the Ministry of Education, Science and Technology (NRF- 2019R1A2C1010851) and by the Korea Astronomy and Space Science Institute grant funded by the Korea government (MSIT; project No. 2024-1-841-00).
We are grateful to the staff of the TRAO who helped to operate the telescope. The TRAO is a facility operated by the KASI. 
\end{acknowledgements}

\appendix


\section{Observing Targets} \label{sec:targets}

Here we describe information about the targets such as their star-forming activities, distances, and the measurements of the LOS and POS magnetic field strengths. 


\subsection{W3OH}

W3OH is one of active star-forming sites in W3 giant molecular cloud (GMC; \citealt{Rivera2013}) which contains a young ultracompact H{\sc ii} (UC H{\sc ii}) region and rich OH masers \citep{Dreher1981}. There is a water maser source named as W3 (H$_2$O) at $\sim$6$''$ east of W3OH where a hot core with a young massive protobinary system is located \citep{Chen2006}. The distance of W3OH is given to be  2.0$_{-0.23}^{+0.29}$ kpc from the measurement of Gaia-DR2 parallaxes of OB stars in W3 GMC \citep{Navarete2019}. 

A CN Zeeman measurement has been made to estimate the LOS magnetic field strength of 1.10 $\pm$ 0.33 mG toward W3OH \citep{Falgarone2008}.  
Dust polarization measurements toward W3OH have been made by the 10 m Heinrich Hertz Telescope and JCMT at 1.3 and 0.85 mm, respectively \citep{Glenn1999, Matthews2009}. However, there is no published result yet on the POS magnetic field strength toward this source.

\subsection{OMC-1}

OMC-1 is located in Orion A molecular cloud at 388 pc $\pm$ 5, which is the nearest high-mass star-forming region \citep{Kounkel2017}. In this region, there are two dense clumps, the Becklin-Neugebauer-Kleinmann-Low (BN/KL) at the northern place and the Orion S at the southern place of the OMC-1 (Figure \ref{fig:dust}; \citealt{Becklin1967, Kleinmann1967, Batria1983, Haschick1989}). The BN/KL clump is known to contain explosive molecular outflow activities which are believed to be made by the interaction of young stars in the core of the clump or a protostellar merger \citep{Gomez2005,Bally2005}. 

Magnetic field strengths in the OMC-1 have been estimated using the Zeeman effect and dust polarization. The LOS magnetic field strengths were measured to be in the range from 0.38 to 16 mG through the Zeeman effect measurements of CN and OH masers (e.g., \citealt{Johnston1989,Falgarone2008,Cohen2006}). The POS magnetic field strengths have been estimated to be from 0.8 to 26.4 mG through the measurements of dust polarization using interferometry and single dish observations (e.g., \citealt{Houde2009, Pattle2017, Chuss2019, Guerra2021, Hwang2021}). The LOS and POS magnetic field strengths in our observed region were measured to be 0.36 and 4.5 mG by the CN Zeeman effect and dust polarization observations using the IRAM and the JCMT, respectively \citep{Falgarone2008, Hwang2021}. The LOS magnetic field strength is given as a mean value of the CN Zeeman measurements within a circular IRAM beam size.
The beam sizes of the IRAM and the JCMT are 23$''$ and 14$''$ at 113 and 353 GHz, respectively.

\subsection{NGC 2024}

NGC 2024 is an active star-forming region in Orion B molecular cloud and associated with a HII region, a massive star cluster, ionizing B stars, and stars at various phases of evolution \citep{Mezger1988, Lada1991, Chandler1996}. The HII region is located in front of the molecular ridge of NGC 2024 \citep{Matthews2002}. 
We adopted the distance of 358.3 $\pm$ 15.0 pc for NGC 2024 which was determined by the proper motions of stellar substructures in the Orion Complex with Gaia EDR3 \citep{Swiggum2021}. 

\citep{Falgarone2008}. The POS magnetic field strength is estimated to be 2.2 mG by using the BIMA at the position of a protostellar source FIR 5, which is selected as the central position of our TRAO observations \citep{Alves2011}. The beam size of the BIMA observations is 2$''$.45 $\times$ 1$''$.48, which is much smaller than that of the CN Zeeman observations. \citet{Matthews2002} made dust polarization observations toward NGC 2024 using the SCUPOL on the JCMT, from which the POS magnetic field strength was measured to be  56 $\mu$G.

\subsection{Mon R2}

Mon R2 is a hub-filament system containing a central massive hub and several filaments converged towards the hub \citep{Rayner2017,Trevino2019,Kumar2022} and located at the distance of 778 $\pm$ 42 pc which is given by the measurement with Gaia DR1 data \citep{Zucker2019}. In the central hub, there are five IRS sources, an ultracompact HII region, and gravitationally bounded multiple cores \citep{Rayner2017}. The central hub shows a rotating flattened structure and is  connected with filaments where longitudinal gas flow exists toward the hub\citep{Trevino2019}.

The measured LOS magnetic field strengths using OH masers are in the range from 0.003 to 0.4 mG \citep{Knapp1976}. There has been no LOS strength measurement using the CN Zeeman effects toward Mon R2. Since the CN line intensity towards Mon R2 is as strong as that of other targets, it is expected that the CN Zeeman effect can be likely detected. \citet{Hwang2021} have obtained the POS magnetic field strengths in Mon R2 with the JCMT observations, which vary from 0.02 to 3.64 mG with a mean value of 1.00 $\pm$ 0.06 mG. 
 
\subsection{DR21}

DR21 is a massive star-forming region located in Cygnus X cloud with a distance of 1.4$^{+0.12}_{-0.11}$ kpc \citep{Schneider2006, Rygl2012}. 
DR21 shows a filamentary ridge structure (e.g., \citealt{Schneider2010}) and contains a massive dense core. DR21(OH) associated with various masers \citep{Zapata2012} is located at peak intensity positions of CN and dust emission distributions shown in Figure \ref{fig:peak_pos}.

The measured LOS magnetic field strength using OH absorption is 0.13 mG  \citep{Koley2021}. \citet{Falgarone2008} have estimated magnetic field strengths of 0.36 and 0.71 mG at two positions of DR21(OH) using the CN Zeeman effects. The POS magnetic field strength is measured to be 0.63 $\pm$ 0.18 mG at the north region of DR21 containing DR21(OH) \citep{Ching2022}.

\subsection{S140}

S140 is located at the molecular cloud L1204 at the edge of the Cepheus ring which is believed to be formed by supernova explosions and stellar winds from the massive stars \citep{Kun1987, Preibisch2002}. S140 contains a diffuse HII region and PDR created by a B0V star (HD 211880) at the edge of the molecular cloud L1204 \citep{Crampton1974}.  \citet{Harvey2012} resolved infrared sources in S140 using the FORCAST camera on the Stratospheric Observatory for Infrared Astronomy (SOFIA). 
The distance of S140 is suggested to be 906 $\pm$ 5 pc with Gaia DR3 data \citep{Szilagyi2023}. 

OH and CN Zeeman observations measured the LOS magnetic field strengths of 2.8 and 0.25 $\pm$ 0.09 mG, respectively \citep{Baudry1997, Falgarone2008}. The POS magnetic field strength for this target was measured to be $\sim$ 0.4 mG by the DCF method using  JCMT observations \citep{Curran2007}. 

\section{Comparison of CN and dust emitting areas} \label{sec:area}

\begin{figure}[htb!]
\epsscale{1.1}
\plotone{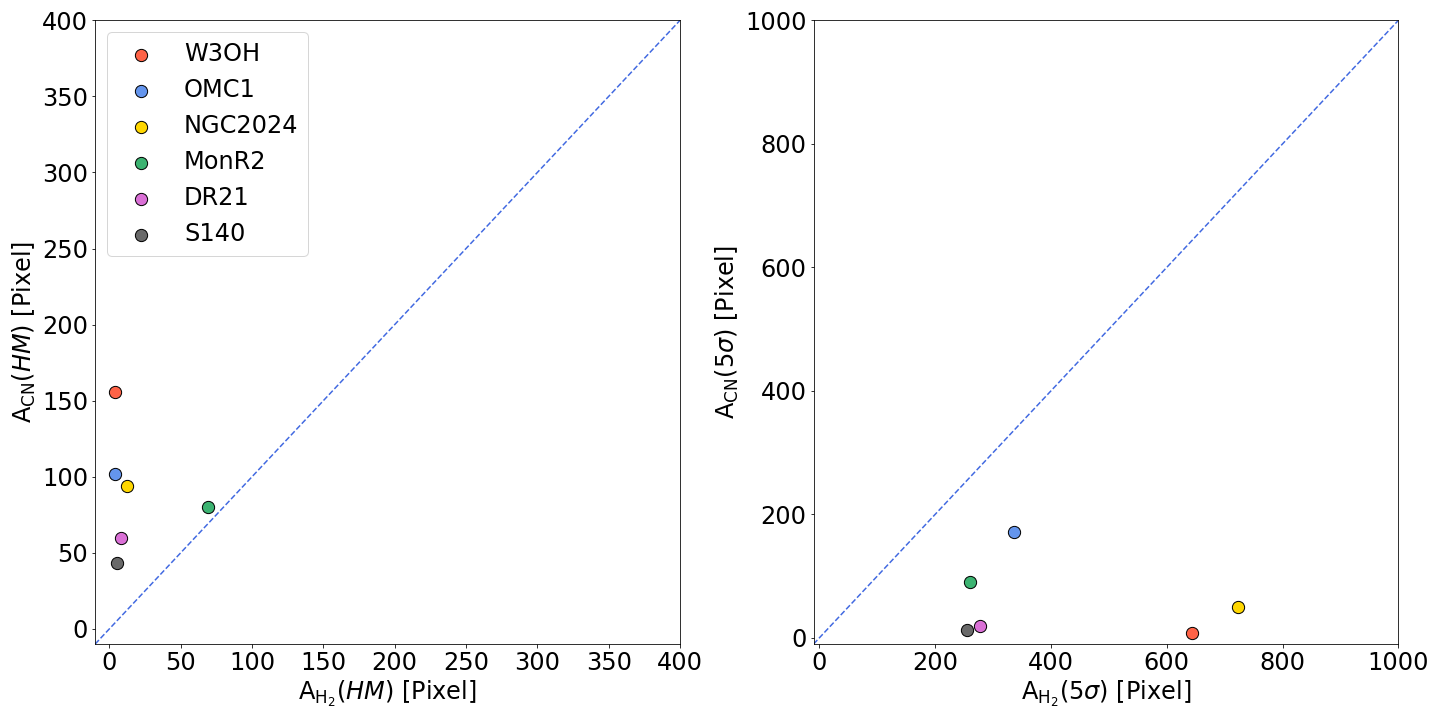}
\caption{Comparison for the areas within HM contours (left) and  $>$5$\sigma$ level contour level (right) in H2 and CN column density distributions for the targets. The blue dashed lines represent the lines where the areas obtained from the CN and H$_2$ column density maps are the same. 
\label{fig:area}}
\end{figure}

Here we compare the CN and dust emitting areas enclosed with two different contour levels for the targets. The one is to use the HM contours for two column densities. The comparison can be made with the areas enclosed with the HM contours for two column density distributions as shown in the left panel of Figure \ref{fig:area}. The HM contour areas of CN column density are at least seven times larger than  than the H$_2$ HM contour areas for all targets, except for Mon R2 for which The HM contour areas of CN column density is comparable to the H$_2$ HM contour areas.
The other one is to use the 5 $\sigma$ contour level for the comparison of low density parts in the targets. As shown in the right panel of Figure \ref{fig:area}, all targets have the area within the H$_2$ column density contour of $>$5 $\sigma$ level much larger than the area of CN column density contour with $>$ 5 $\sigma$ level. 

\end{document}